\documentclass[10pt]{article}
\usepackage{emulateapj-djb}
\usepackage{epsfig}
\def\ci     {C{\sc i}}
\def\h2o    {\ifmmode{1_{10}\leftarrow 1_{01}}\else
            {{$1_{10}\leftarrow 1_{01}$ }}\fi}
\def\kms    {\ifmmode{{\rm ~km~s}^{-1}}\else{~km~s$^{-1}$}\fi}
\def\cm     {\ifmmode{{\rm ~cm}^{-2}}\else{~cm$^{-2}$}\fi}
\def\nh2    {\ifmmode{~N({\rm H}_2)}\else{~$N$(H$_2$)}\fi}
\def\43     {\ifmmode{4 \leftarrow 3 }\else{4$\leftarrow$3 }\fi}
\def\54     {\ifmmode{5 \leftarrow 4 }\else{5$\leftarrow$4 }\fi}
\submitted{To appear in Astrophysical Journal Letters}

\begin{document}
\title{Redshifted molecular 
absorption systems towards PKS~1830$-$211 and B~0218+357:
submillimeter CO, \ci~and H$_2$O data}

\author{Maryvonne Gerin\altaffilmark{1} (gerin@ensapa.ens.fr),
Thomas G. Phillips\altaffilmark{2} (phillips@tacos.caltech.edu),
Dominic J. Benford\altaffilmark{2} (dbenford@tacos.caltech.edu),
Ken H. Young\altaffilmark{2,3} (rtm@cfacrm.harvard.edu), 
Karl M. Menten\altaffilmark{4} (menten@cfa.harvard.edu),
Brenda Frye\altaffilmark{5} (bfrye@astron.berkeley.edu)}

\altaffiltext{1}{URA336 of CNRS and Radioastronomie Millim\'etrique,
Laboratoire de Physique de l'ENS, 24 Rue Lhomond, 75231 Paris cedex 05, France}
\altaffiltext{2}{California Institute of Technology, Pasadena, CA 91125, USA.}
\altaffiltext{3}{Harvard-Smithsonian Center for Astrophysics, 60 Garden Street,
Cambridge, MA 02138, USA.}
\altaffiltext{4}{Max Planck Institute f\"ur Radioastronomie, Auf dem H\"ugel
69, D-53121 Bonn, Germany}
\altaffiltext{5}{Radio Astronomy Laboratory and Department of Astronomy,
University of California, Berkeley, CA 94720, USA.}

\begin{abstract}
  
  We have detected the $J=\43 $ rotational transition of $^{12}$CO in
  absorption at $z$~=~0.89 towards the quasar PKS~1830$-$211, but not the
  $^{12}$CO ($\54 $) or the $^3P_1 \leftarrow ^3P_0$ fine structure
  line of neutral carbon. The intervening molecular medium thus has a
  total $^{12}$CO column density of $10^{18} \cm \leq N({\rm CO}) \leq
  5 \times 10^{18}$\cm with a most likely value of $N$(CO)~$\simeq
  2\times 10^{18} \cm$, which corresponds to the large column density
  of molecular hydrogen of \nh2 = $2.5 \times 10^{22}$ \cm~ and a
  reddening of A$_v$~=~25 magnitudes.  The $^{12}$CO excitation
  temperature is low, below 15 K. Comparison with the molecular
  absorption results of Wiklind and Combes (\cite{wkc}) shows that the
  absorbing material has similar molecular abundances to Galactic dark
  clouds. We find an upper limit for atomic carbon of $N$(\ci)$\leq
  10^{18}\cm$, which again would be the case for most Galactic dark
  clouds.
  
  We also report new observations of the absorbing system towards
  B~0218+357 at $z$~=~0.68. We have tentatively detected the $^{13}$CO
  ~($\43 $) line, but for H$_2$O, although a feature is seen at the
  correct velocity, due to the inadequate signal to noise ratio we
  report only an upper limit for the fundamental line of ortho water
  vapor. The tentative detection of the $^{13}$CO $J=\43 $ line
  implies that the $^{13}$CO excitation temperature is lower than 20 K
  and the column density is fairly large, $4\times 10^{16}\cm \leq$
  $N$($^{13}$CO) $\leq 2.2\times 10^{17}$ \cm, with a likely value of
  $N$($^{13}$CO) $\simeq 10^{17}$ \cm, giving rise to saturated
  absorption in the ($J=2\leftarrow1$) transition. The total column
  density of molecular gas is again large in this source, \nh2 $\geq
  2\times10^{22}$ \cm, which corresponds to a reddening larger than 20
  magnitudes.
 
\end{abstract}

\keywords{(Galaxies: quasars:) absorption lines ---
Galaxies: quasars: individual (PKS1830$-$211,B0218+357) --- Galaxies: ISM ---
ISM: Molecules}

\setlength{\textfloatsep}{12pt}

\section{Introduction}

\begin{figure*}[b]
\center{{\bf Table 1.}

Flux densities and line intensities from PKS~1830$-$211.

\bigskip

\begin{tabular}{|c|c|cc|cc|c|} 
\hline \hline
Line & Observed& \multicolumn{2}{|c|}{March 1996} & 
\multicolumn{2}{|c|}{July 1996} & Combined \\
 & Frequency & Continuum  & Line  & Continuum & Line
  & Line \\
\hline
$^{12}$CO($\43 $) & 244 GHz & $0.82 \pm 0.05$ & $0.5 \pm 0.18$ & 
$0.36 \pm 0.05$ & $0.6 \pm 0.25$ & $0.5 \pm 0.14$ \\
\ci($1\leftarrow 0$) & 261 GHz & $0.75 \pm 0.05$ & $\leq 0.2$ & & & \\
$^{12}$CO($\54 $) & 305 GHz & $0.61 \pm 0.03$ & $\leq 0.3 $ & 
$0.27 \pm 0.03$ & $\leq 0.4 $ & $\leq 0.2$ \\
\hline
\end{tabular} 

{\small \parbox{5.5in}
{The continuum fluxes are in Jy, while the line intensities are the
fractional absorption depths given relative to the SSB continuum level
with a velocity resolution of 6 \kms. The line width of the CO($\43 $)
line is $18 \pm 3$ \kms.  Upper limits and errors are given at the
1$\sigma$ level.}}
}
\end{figure*}

\begin{figure*}[t]
\begin{center}
{\bf Table 2.}

Flux densities and line intensities from B~0218+357.\bigskip

\begin{tabular}{|c|c|cc|cc|} 
\hline \hline
Line & Observed & \multicolumn{2}{|c|}{September 1995} & 
\multicolumn{2}{|c|}{October 1996} \\
 & Frequency & Continuum & Line & Continuum & Line \\
\hline
$^{13}$CO($\43 $) & 261 GHz & & & $0.21 \pm 0.02$ & $0.6 \pm 0.2$ \\
H$_2$O \h2o & 330 GHz  & $0.18 \pm 0.02$ & $\leq  0.5 $ & & \\
\hline
\end{tabular}

{\small\parbox{4.5in}
{The continuum fluxes are in Jy, while the line absorption intensities
are given relative to the SSB continuum level with a velocity
resolution of 1.5 \kms ~for the $^{13}$CO line and 6 \kms ~for the
H$_2$O line.  Upper limits are given at the 1$\sigma$ level.}}
\end{center}  
\end{figure*}

\subsection{PKS~1830$-$211}

In addition to emission line observations, the interstellar medium of
distant galaxies can be probed by absorption lines when a background
continuum source is present. The intervening galaxies of
gravitationally lensed QSOs fulfill this condition. Wiklind and Combes
(\cite{wka}, \cite{wkb}) have searched for molecular line absorption
towards such sources and in a few cases have detected molecules such
as HCO$^+$, HCN, HNC, and CO that are commonly found in dark
interstellar clouds within the Galaxy.  The most remarkable example is
the line of sight toward the radio source PKS~1830$-$211, for which the
redshift of the intervening galaxy was determined by molecular
absorption line spectroscopy to be $z$~=~0.88582 (Wiklind \& Combes
\cite{wkc}).

The flat spectrum radio source PKS~1830$-$211 is well established as
being a gravitationally lensed object (Rao \& Subrahmanyan \cite{rao})
with no optical counterpart (Djorgovsky et al.  \cite{djorgovsky}).
Recent radio images (van Ommen et al.  \cite{ommen}) show two major
components separated by about 1 arcsec, with a minor, possibly
demagnified, third component between them. The system has been modeled
as two images of a background source with a core-jet morphology,
magnified and distorted by an intervening galaxy which partially
obscures the southwest image. There is a time delay between the two
lensed images of $44\pm 9$ days (van Ommen et al.  \cite{ommen}).  The
flux density varies with time, between 0.5 and 1.3 Jy at 1.3 mm
(Steppe et al. \cite{hs92}, \cite{hs93}), with the SW image containing
between 40 and 45\% of the total flux.

Wiklind and Combes (\cite{wkc}) have detected rotational lines of HCN,
HNC, HCO$^+$, H$^{13}$CO$^+$, CS, and N$_2$H$^+$ in the intervening
galaxy. The lines are moderately narrow with a maximum linewidth of 30
\kms. Though saturated, the absorptions do not reach the zero level,
indicating that the absorbing system covers only $\sim$36\% of the
continuum source. This is verified by the recent maps made with the
BIMA interferometer at 3mm (Frye, Welch, \& Broadhurst \cite{frye})
where molecular absorptions are detected towards the SW image only.

The data set of Wiklind and Combes (\cite{wkc}) does not include CO
lines, because at the redshift of 0.89, the three lowest CO
transitions are shifted to frequencies suffering severe atmospheric
absorption and are impossible or difficult to observe from the ground.
The next two rotational lines, however, can be observed: the \43 line
at a rest frequency of 461.040 GHz and the \54 line at a rest
frequency of 576.268 GHz are redshifted to 244.477 GHz and 305.579
GHz, respectively.  In this paper we report observations at these two
line frequencies as well as of the $^3P_1 \leftarrow ^3P_0$ fine
structure line of neutral carbon at 492.161 GHz, redshifted to 260.980
GHz.

\subsection{B~0218+357}

H{\sc I} absorption has been detected at $z$~=~0.68466 toward the
Einstein ring B~0218+357 (Carilli, Rupen, \& Yanny \cite{carilli}),
which is the prototype of a lensed object where the lensing galaxy is
perfectly aligned with the QSO. The column density of absorbing
material is very high, as shown by the presence of saturated lines of
HCN, CO, and its isotopes (Wiklind \& Combes \cite{wkb}). Formaldehyde
has also been detected at centimeter wavelengths (Menten \& Reid
\cite{menten}), as a weak absorption feature.  The small linewidth of
the molecular absorption, 5 -- 10 \kms, indicates that we are dealing
with a single molecular cloud or a small number of clouds along the
line of sight.  Estimates of the molecular hydrogen column density
range from $3\times 10^{22}\cm$ (Menten \& Reid \cite{menten}) to
$5\times 10^{23}\cm$ (Combes \& Wiklind \cite{cw95}).  To aid in the
understanding of the physical nature of the absorbing cloud we have
searched for the $^{13}$CO(4$\leftarrow$3) transition.

Combes and Wiklind (\cite{cw95}) searched unsuccessfully
for molecular oxygen.  We have searched for the fundamental transition
of ortho water vapor at 556.936 GHz towards this object, redshifted to
330.592 GHz.

\section{Observations}

We observed PKS~1830$-$211 using the Caltech Submillimeter Observatory
(CSO) 10.4m telescope in March and July of 1996, under good weather
conditions. The observed position was taken from NED as
R.A.(1950)=18:30:40.600, Dec(1950)=$-$21:06:00.0, near the location of
the modeled lensing galaxy (Subrahmanyan et al. \cite{sub90}). The
telescope is equipped with SIS receivers operated in double sideband
(DSB) mode.  The observations were performed by beam switching using
the chopping secondary with a frequency of 1 Hz and a beam separation
of 90$''$.  The pointing was checked on Jupiter, IRC~10216, and CIT~6
and was found to be accurate to about 3$''$ rms.  The frequency
calibration was checked on G~34.3+0.2, a nearby ultracompact HII
region.  We searched for the CO($\43 $) line redshifted to 244.477
GHz, the CO($\54 $) line at 305.579 GHz and the \ci~ ($^3P_1
\leftarrow ^3P_0$) line at 260.980 GHz.  Single sideband (SSB) system
temperatures, relevant to the line observations, were between 200 and
300 K at 244 and 261 GHz and $\simeq 500$ K at 305 GHz.
 
The spectra were simultaneously analysed with two acousto-optic
spectrometers (AOS), one with a total bandwidth of 500 MHz and an
effective spectral resolution of 1.5 MHz and the other with 1.5 GHz
total bandwidth, 0.9 GHz of which were usable due to the limited IF
bandpass of the receiver.  For the \ci~ observations the 500 MHz wide
AOS was used.  The angular resolution of the CSO is 28$''$ at 244 GHz,
26$''$ at 261 GHz, and 23$''$ at 305 GHz. The conversion factors from
Kelvins to Janskys are obtained from efficiency values given in the
CSO Observing Manual and are 43 Jy/K at 230 GHz and 46 Jy/K at 345
GHz.

B~0218+357 was observed in 1995 September for two nights and in 1996
October for 4 nights. The source position, again obtained from NED, is
R.A.(1950)=02:18:04.1, Dec(1950)=35:42:32.  In addition to the 500 MHz
AOSs, we used a narrow band AOS with a total bandwidth of 50 MHz and
an effective spectral resolution 100 kHz to identify possible narrow
features.

For both sources, the spectra are presented in the LSR velocity scale.
To obtain heliocentric velocities, subtract 12 \kms~for the
observations toward PKS~1830$-$211 or add 2 \kms~in the case of
B~0218+357.

\section{Results}
 
\begin{figure*}[b]
\centerline{\epsfig{file=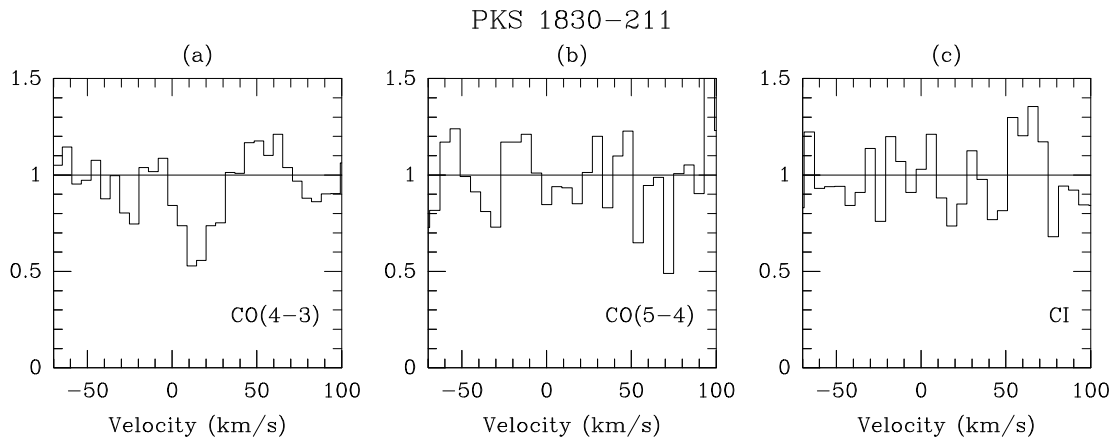,width=6.0in}}
\label{spectra}
\figcaption{\small{SSB spectra normalized to the continuum level towards
PKS~1830$-$211 for (a) CO($\43 $) at 461.040 GHz redshifted to
244.477 GHz, (b) CO($\54 $) at 576.268 GHz redshifted to
305.579 GHz, (c) \ci($1\leftarrow 0$) at 492.161 GHz redshifted to
260.980 GHz.}}
\end{figure*}

\subsection{Continuum Flux}

The continuum emission of PKS~1830$-$211 was readily detected at all
three frequencies in both backends.  The results are summarized in
Table~1 for both observing runs.  The spectral index in the millimeter
and submillimeter range is about -1.3.  The continuum level was
fainter by a factor of about 2 in July 1996; this reduction of the
flux density is in agreement with the measurements at lower
frequencies made at a similar epoch with the IRAM and SEST telescopes
(F. Combes, private communication).  Also shown in Table~1 are the
line intensities relative to the SSB continuum level.  While it is
possible that the relative contributions of the two images to the
total continuum may change with time, so causing the observed
fractional absorption of the continuum to vary, this apparently is not
a particularly significant factor in this experiment, since the
fractional absorption of the CO ($4\leftarrow3$) is the same within
the error bars for the two epochs. Thus it seems reasonable to combine
the results, at least in the case of limited signal to noise ratio, as
presented here.  The final column of Table~1 shows the combined
absorption fraction.

For B~0218+357, we measured a continuum flux density of 0.18 Jy at 330
GHz in 1995 September and 0.21 Jy at 261 GHz in 1996 October. Table~2
presents the B~0218+357 flux measurements and line intensities in the
same format as Table~1.

\subsection{Absorption lines toward PKS~1830$-$211}

Figure~1 
shows the combined single-sideband (SSB) spectra
of the two sessions normalized to the observed continuum level for
each session, for CO($\43 $), CO($\54 $), and \ci.  Only the CO($\43 $)
absorption line was detected. The absorption reaches the same depth as
the HCN and HCO$^+$ lines detected by Wiklind and Combes (\cite{wkc}),
about 40\% of the SSB continuum level.  Following Wiklind and Combes'
(\cite{wkc}) interpretation of their HCN and HCO$^+$ observations, we
conclude that only part of the QSO continuum emission is being
absorbed, and that the CO ($\43 $) line is, in fact, optically thick.
This is confirmed by BIMA measurements of Frye et al. (\cite{frye})
who show that only one of the lensed QSO images has a line of sight
that passes through the absorber.  Therefore, the CO($\43 $) absorption
is also probably saturated and has a similar covering factor to the
HCN and HCO$^+$ absorptions.  The line appears at V$_{LSR}$ = 15 \kms,
hence V$_{hel}$~=~3~\kms, in agreement with the redshift given by
Wiklind and Combes (\cite{wkc}).  The linewidth is 18 \kms, somewhat
smaller than the linewidth of 30 \kms~given for the most saturated
HCO$^+$ line.  The difference in linewidth between CO and HCO$^+$ is
likely to be due to differences in line opacities since the CO($\54 $) line is
not detected at a similar sensitivity.  We conclude from our data that
the opacity of the CO($\43 $) line is larger than 3 in order to
get a saturated absorption, while the opacity of
the CO($\54 $) line is smaller than 1 at the 1$\sigma$ level.

\begin{figure*}[t]
\centerline{\epsfig{file=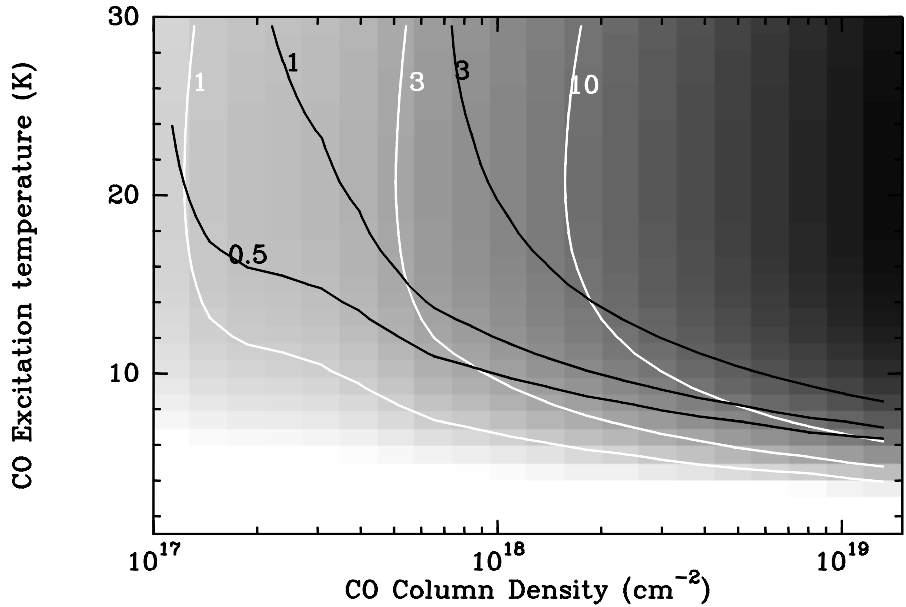, width=5.0in}}
\figcaption{\small{Opacity of the CO($\43 $) (grey scale and
white contours) and CO($\54 $) (black contours) lines as functions
of the total CO column density and excitation temperature.  The
allowed parameter space lies above the white curve labeled 3 and
below the black curve labeled 1.}}
\label{opacity}
\end{figure*}

\begin{figure*}[b]
\centerline{\epsfig{file=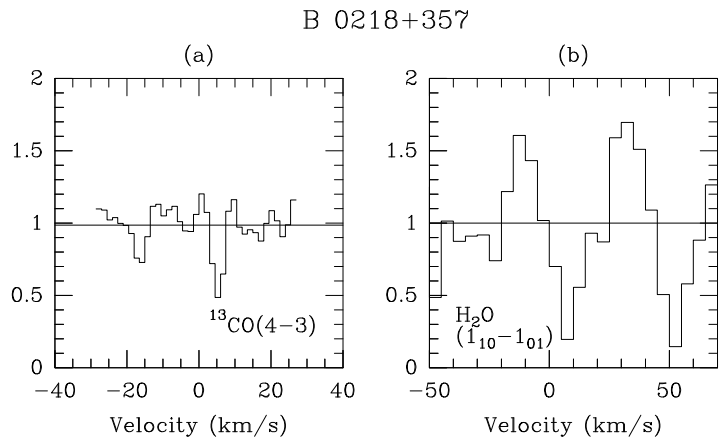,width=6.0in}}
\figcaption{\small{SSB spectra normalized to the continuum level towards
B~0218+357 for (a) $^{13}$CO($\43 $) at 440.765 GHz redshifted to
261.634 GHz and (b) H$_2$O (\h2o ) at 556.936 GHz redshifted to
330.592 GHz.}}
\label{spectra2}
\end{figure*}

The detection of CO($\43 $) absorption with a moderately large opacity
combined with the non-detection of the CO($\54 $) line puts
constraints on the CO excitation temperature.  We show in
Figure~2 a comparison of the opacities of the ($\43 $) and
($\54 $) lines for a range of excitation temperatures and column
densities.  These calculations have been done assuming LTE.  The
excitation temperature of the gas must be smaller than 15 K. Assuming
T$_{ex}$ = 8 K, the upper limit for the excitation temperature of
HCO$^+$, we obtain the following range for the CO column density :
$10^{18} \cm \leq N({\rm CO}) \leq 5\times10^{18} \cm$ with a most
likely value of 2$\times 10^{18}$ cm$^{-2}$.

With a $^{12}$CO abundance relative to H$_2$ of 8$\times 10^{-5}$, as
in Galactic dark clouds (Irvine, Goldsmith, \& Hjalmarson
\cite{irvine}), the H$_2$ column density is 2.5$\times 10^{22}$
cm$^{-2}$.  These CO and H$_2$ column densities are in good agreement
with the estimates proposed by Wiklind and Combes (\cite{wkc}) and
show that the molecular gas seen in absorption has similar molecular
abundances to that of Galactic dark clouds.

We have not detected \ci~ absorption, with a 1$\sigma$ noise level of 20\%
of the continuum flux in 6 \kms~channels. This means that the
opacity of the \ci~ line must be of the order of 1 or lower.  For \ci~
at 8 K, we get $N$(\ci) $\leq$ 10$^{18}$ cm$^{-2}$, thus [C]/[CO] $\leq$
1, which again is in accord with Galactic dark clouds for which
[C]/[CO] = 0.1 - 1.0 (Keene \cite{keene}).

\subsection{The line of sight towards B~0218+357}

Although an absorption feature appears at the correct velocity, we
cannot report the detection of the redshifted H$_2$O ($\h2o $) line
due to inadequate signal to noise ratio, but we did obtain a tentative
detection of the $^{13}$CO($\43 $) line toward B~0218+357. The spectra
are shown in Figure~\ref{spectra2}.  The line width of the
$^{13}$CO($\43 $) line is small compared to the lower $J$ lines,
suggesting an opacity of the order of 1, more precisely $0.5 \leq
\tau_{\43 } \leq 3$.  However, the $^{13}$CO($2\leftarrow 1$) line is
certainly saturated with an opacity larger than 5 since the
C$^{18}$O($2\leftarrow 1$) line is nearly saturated (Combes \& Wiklind
\cite{cw95}). Using similar calculations as for PKS~1830$-$211 and
assuming a Doppler linewidth of 5 \kms, we conclude that the
excitation temperature of the $^{13}$CO gas is lower than 20 K, with a
likely value in the range of $\simeq 12 $ K.  Then, the column density
lies in the range $ 4 \times10^{16} \cm \leq$
$N$($^{13}$CO)~$\leq$~2.2$\times10^{17} \cm$.  These data can be used to
set a lower limit on the molecular gas column density. We use a
$^{13}$CO abundance relative to H$_2$ of $2 \times 10^{-6}$, which
lies at the upper end of the derived range for dark clouds (Irvine et
al. \cite {irvine}) and get $N$(H$_2$) $\geq$ $2 \times 10^{22} \cm$.
Finally, it is well known from Galactic molecular cloud studies that
the LTE approximation is usually reasonable, particularly in the case
of CO molecules in dark clouds.  In the case here, an LVG calculation
with $N$($^{13}$CO) = 10$^{17} \cm$~ and a linewidth of 5 \kms~shows
that the maximum kinetic temperature compatible with the observations
is 50 K for an H$_2$ density of 500 cm$^{-3}$, decreasing to 15 K for
$N$(H$_2$)~=~10$^4$ cm$^{-3}$.  In Galactic dark clouds it is the high
density, low kinetic temperature case which is observed
(Falgarone \& Phillips \cite{FandP}), essentially the LTE result.

\section{Conclusion}

We have made new observations on the line-of-sight towards
PKS~1830$-$211 and B~0218+357. We show that the gas is cold in both
lines of sight, with T$_{ex}$ $\leq$ 15 K and $\leq$ 20 K
respectively.  We have determined the gas column density in both
cases, finding $\nh2 \simeq 2.5\times10^{22} \cm$ for PKS~1830$-$211 and
$\nh2 = 2-5\times 10^{22} \cm$ for B~0218+357.  We find the gas along
the line of sight to PKS~1830$-$211 to be chemically typical of a
Galactic dark cloud.

There are various ways to enhance the quality of the description of
the physical conditions and characteristics of the absorbing cloud. It
should be possible, for example, to search for specific tracers of the
chemistry and the elemental gas phase abundances, such as CCH, CS, SO,
and DCN, because their relative abundances are expected to be
sensitive to the physical conditions. Deuterated species, though
difficult to observe, are particularly interesting since the
efficiency of the deuterium fractionation mechanism differs
considerably in the high ionization and low ionization chemical phases
of dark interstellar clouds (Gerin et al.  \cite{gerin}).

\bigskip

Work at the CSO is supported by NSF grant \#AST96-15025.

\end{document}